\documentclass[twocolumn,prl,aps,amsmath,superscriptaddress,showpacs]{revtex4}

\usepackage{epsfig}
\usepackage{float}

\def\dk         {\frac{d\,{\bf k}}{\(2 \pi\)^3}}
\def\dk1        {\frac{d\,{\bf k}_1}{\(2 \pi\)^3}}

\def\ga         {\alpha}

\def\gd         {\delta} 
\def\gee        {\epsilon}

\def\go         {\omega}
\def\gr         {\rho}

\def\la         {\langle}
\def\ra         {\rangle}
\def\kk         {{\bf k}}
\def\qq         {{\bf q}}
\def\QQ         {{\bf Q}}
\def\rr         {{\bf r}}

\def\GG         {{\bf G}}

\def\tGa        {\widetilde{\Gamma}}

\def\tChi       {\widetilde{\chi}}

\def\ff         {{\bf f}}

\def\fxc        {\ff_{xc}\(\go\)}

\renewcommand{\[}{\left[}
\renewcommand{\]}{\right]}
\renewcommand{\(}{\left(}
\renewcommand{\)}{\right)}

\restylefloat{figure}

\begin{document}
%
%%%%%%%%%%%%%%%%%%%%%%%%%%%%%%%%%%%%%%%%%%%%%%%%%%%%%%%%%%%%%%%%%%%%%%%%%%%%
\title{Electron linewidths of wide--gap insulators: excitonic effects in LiF}
\author{Andrea Marini}
\affiliation{
Departamento de F\'\i sica de Materiales,
Facultad de Ciencias Qu\'\i micas,
Universidad del Pais Vasco, 
Centro Mixto USPV/EHU--CSIC and  Donostia International Physics Center.
E--20018 San Sebasti\'an, Basque Country, Spain}
\author{Angel Rubio}
\affiliation{
Departamento de F\'\i sica de Materiales,
Facultad de Ciencias Qu\'\i micas,
Universidad del Pais Vasco, 
Centro Mixto USPV/EHU--CSIC and  Donostia International Physics Center.
E--20018 San Sebasti\'an, Basque Country, Spain}
\date{\today}

\begin{abstract} 
Based on a recent exchange--correlation kernel developed 
within Time--Dependent--Density--Functional Theory  we derive
a practical and general expression for the three--point vertex function.
We show that excitonic effects in 
LiF strongly modifies the low--energy electron linewidths leading to
linear scaling with quasiparticle energy. 
We also prove that, in contrast to previous results for the 
electron gas, simple metals and
semiconductors, vertex corrections in the self--energy and in the screening
function do not compensate each other.
\end{abstract} 
\maketitle

The experimental quasiparticle band structure of
bulk metal and semiconductor systems has been
successfully explained by the
$GW$ self--energy scheme~\cite{hedin,gwrep,rmp} in
its simplest non self--consistent $G_0W_0$ implementation. 
Similarly, the quasiparticle linewidths of simple and noble metals
have been studied extensively~\cite{pedro}, but
a first--principles description of the electronic contribution to the
electron/hole linewidths in semiconductor and insulators is not yet available.
The reason is that
the low--energy quasiparticle dynamics in semiconductors
tends to be dominated by inelastic phonon scattering, 
the electronic contribution playing a minor role.
However this scenario changes drastically when the quasiparticle energy
is larger than the minimum energy required to excite an
electron--hole pair. Above this threshold (that is zero in metals and 
approximatively twice the band gap in insulators) the rapidly
increase in density of electron--hole pairs dominates the
quasiparticle damping, resulting in a mainly electronic contribution to the lifetime.
It is well known that in insulators, at difference with metals, 
the attractive interaction between electrons and holes can lead to
the formation of a bosonic--like excitonic state~\cite{rmp}.
Excitons modify remarkably the optical and energy-loss
spectra and, consequently,
the microscopical mechanisms responsible for the quasiparticle
damping. This effect is stronger in wide--gap insulators like LiF.

In this communication we tackle the problem of evaluating the
impact of the excitonic effects on the quasiparticle\,(QP) dynamics of LiF,
using a simplified vertex function in the electronic self--energy.
An efficient approximation for the {\it three--point} 
many--body vertex function is given in terms of the 
{\it two--point} exchange--correlation kernel
$f_{xc}$~\cite{noi}, recently developed in the framework of 
Time--Dependent--Density--Functional Theory\,(TDDFT)~\cite{rmp}.
As result we show that the electronic linewidths of LiF
display a linear dependence as function of the QP energy, 
that can be traced back to the incipient excitonic effects 
induced by $f_{xc}$~\cite{metals}.
Understanding the inelastic mechanism which dominates the phase
coherence time is crucial to the field of quantum transport in mesoscopic
and nanostructred materials. Thus this work
is the first steps toward a full first--principles description of the quasiparticle
dynamics of semiconductor and insulators.

\begin{figure}[H]
\begin{center}
\epsfig{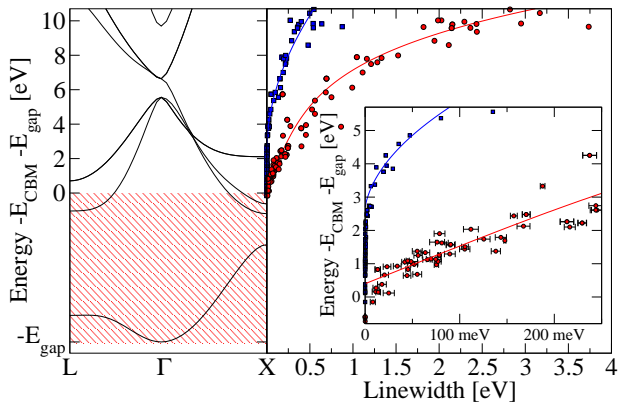}
\end{center}
\vspace{-.5cm}
\caption{
\footnotesize{(Color on-line) 
Left panel: calculated DFT band--structure of LiF  
(here $E_{CBM}$ and $E_{gap}$ stand for the DFT conduction band
minimum energy and the energy gap).
Right panel: Electron linewidths calculated
``on mass--shell"  as function of
the single--particle energy.
Boxes: RPA $G_0W_0$. Circles: TDDFT--based vertex correction
to the self--energy, i.e, a $G_0W\tGa^{\(1\)}_{TDDFT}$ approach
that turn out to be
very close to a simpler $G_0W$ calculation\,(see text). The dashed area denotes
the forbidden energy region for quasiparticle decay into electron--hole pairs.
Error bars represent the theoretical uncertain due to the
"zero--broadening" extrapolation~\cite{details}.
}}
\label{fig1}
\end{figure}

In the usual one--shot $G_0W_0$ self--energy scheme, it is assumed that a basic 
Density--Functional--Theory\,(DFT) calculation~\cite{dftdetails}
provides good approximation for QP wavefunctions and 
electronic screening (dominated by collective excitations, plasmons
build from independent electron--hole transitions, i.e. 
excitonic effects in the screened Coulomb potential $W_0$ are neglected).
The QP lifetime $\tau_i$ can be calculated with the Fermi golden--rule,
using this non interacting $W_0$ as scattering potential:
$\tau_i^{-1} = -2 \sum_f \left|\Omega_{if}\right|^2 Im \[W_0\(E_i-E_f\)\]$,
where $|i\ra$, $|f\ra$ are the initial and final state, with 
$W_0$ matrix elements $\Omega_{if}$, and energies
$E_i$ $E_f$ such that $E_i-E_f>0$.
This scattering scheme, also known as ``on--mass shell" approximation to the
$G_0W_0$ linewidths, provided
valuable insight into the electron/hole linewidths of 
metals~\cite{pedro}. Therefore, in
Fig.~\ref{fig1} we estimate the electronic line widths of LiF\,(boxes) within
this approximation.
As $W_0$ is calculated in terms of non--interacting 
electron--hole pairs $\tau_i^{-1}=0$ when $E_i-E_f<E_{gap}$ 
($E_{gap}$ the DFT gap):
quasiparticle states with energy $E_i<2 E_{gap}$ 
have zero line width (infinite lifetime).
These states are indicated by the dashed area in Fig.~\ref{fig1}.
Above this region a quadratic
energy dependence of the line width is recovered, as in metals~\cite{smith}.

\begin{figure}[H]
\begin{center}
\epsfig{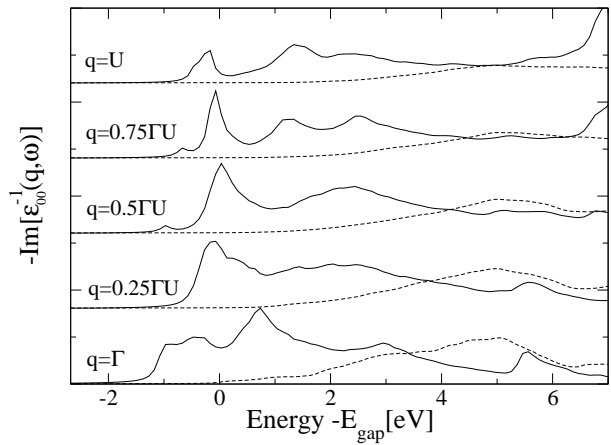}
\end{center}
\vspace{-.5cm}
\caption{
\footnotesize{Calculated loss function of LiF 
for momentum transfer $\qq$ along the direction $\Gamma\,U$. 
Continuous line: TDDFT calculation with an $f_{xc}$ kernel
that mimics excitonic effects~\cite{noi}. Dashed line: RPA.}}
\label{fig2}
\end{figure}

As the short--range screened Coulomb repulsion modifies
drastically the polarization function in LiF,
one is tempted to apply the previous ``on--mass shell"
scheme to analyze the role of excitonic effects on the quasiparticle
dynamics. 
This would correspond to replace 
$W_0$ by the screened coulomb potential $W$ obtained from
the many--body Bethe--Salpeter equation\,(BSE)~\cite{rmp,strinati,dbse}.
In practice, the BSE
sums all the possible binary collisions between electrons and holes,
providing a consistent and successful framework for
the calculation of the interacting polarization function. 
However, the BSE is  computationally very demanding and it becomes
unpractical when the microscopical dielectric matrix $\hat{\gee}(\qq,\go)$
must be calculated for a large set of transfer momenta $\qq$ and
frequencies $\go$,
as it is the case for the calculation of linewidths~\cite{pedro}.
To bypass this difficulty we compute $\hat{\gee}(\qq,\go)$ within a TDDFT framework, 
using an $f_{xc}$
kernel~\cite{noi} that mimics well the  BSE results~\cite{details}.
This performance is illustrated in Fig.~\ref{fig2} for the loss
function $\hat{\gee}^{-1}(\qq,\go)$, that is the relevant quantity to
build the screened Coulomb potential $W$. 
From Ref.~\cite{noi} we know that TDDFT reproduces the experimental loss function,
therefore comparing the TDDFT and random phase approximation\,(RPA) results
of Fig.~\ref{fig2} we see that RPA misses the 
strong weight of the loss spectra just above the band--gap. 
Consequently the inclusion of excitonic effects in this $G_0W$ calculation 
translates into a drastic change of 
the quasiparticle decaying rates\,(red circles in Fig.~\ref{fig1}) compared
to the RPA results\,(blue boxes).

This simple scattering approach, though appealing, 
lacks of theoretical consistency. Exchange--correlation effects
have been included {\it only} in the polarization function, while,
in the spirit of the original work of Hedin~\cite{hedin}, they should be included
in the self--energy as well.
However we will show below that the results obtained within a proper
treatment of 
self--energy and polarization effects do not deviate appreciably from the previous $G_0W$ results.
We start the derivation from the definition of the 
self--energy operator $\Sigma\(1,2\)$, given by~\cite{strinati}
\begin{align}
\Sigma\(1,2\)=i\,\int\,d34\,W\(1^+,3\)G\(1,4\) \tGa\(4,2;3\).
\label{dyson}
\end{align}
Here $G(1,2)$ is the interacting Green's function and $\tGa\(1,2;3\)$ 
the irreducible vertex function (numbers stands for
space, time and spin coordinates). 
The screened Coulomb interaction $W$ is: $W\(1,2\)= v\(1,2\) +
 \int\,d34\, v\(1,3\) \tilde{\chi}\(3,4\) W\(4,2\)$ where $v\(1,2\)$ is
the bare Coulomb interaction and $\tilde{\chi}$ the irreducible
polarization function:
\begin{align}
\tilde{\chi}\(1,2\)=-i\int\,d34\,G\(1,3\)G\(4,1\)
 \tGa\(3,4;2\).
\label{chitilde}
\end{align}
Thus, given an approximation for $\tGa$ the self--energy
is completely defined trough Eqs.\,(\ref{dyson}--\ref{chitilde}) plus the Dyson equation
for $G$.
Electron--hole effects are embodied in the vertex function $\tGa$ that
appears in the self--energy directly, in Eq.\,(\ref{dyson}), and
trough the polarization function, Eq.\,(\ref{chitilde}).
The interplay between those two effects has been strongly debated
in the last years, using different approximations for $\tGa$, and
different levels of self--consistency in the solution of Dyson equation.
However all the systems analyzed in the past are characterized by 
moderate, if not absent, excitonic effects in the polarization function.
Thus even if the use of two--point DFT--based~\cite{mahan,carlolaf,rodolfo} 
or finite order vertex functions $\tGa$~\cite{1ord} can be justified in 
the case of the homogeneous electron gas or 
simple semiconductors, they will be
inadequate in the case of wide--gap insulators\,(e.g. LiF), as well as in the case of other 
strongly correlated systems.
Next we derive a TDDFT approximation to the vertex
function $\tGa$ following the spirit of Ref.~\cite{noi}:
to reproduce the diagrammatic expansion of $\tGa$ obtained within
Many--body perturbation theory.
In the non self--consistent scheme the BSE expresses $\tGa$  in terms
of the independent particle $W_0$\,(calculated using Eq.\,(\ref{chitilde}) 
assuming $\tGa\(1,2;3\)=\gd\(1,2\)\gd\(1,3\)$) 
and the bare DFT Green's function $G_0$ as:
\begin{multline}
\tGa\(1,2;3\)=
 \gd\(1,2\)\gd\(1,3\)+\\ i W_{0}\(1,2\) \int\,d67\, G_0\(1,6\)G_0\(7,2\)
  \tGa\(6,7;3\).
\label{bse}
\end{multline}
When this vertex $\tGa$ is inserted in Eq.\,(\ref{chitilde}), the corresponding
equation for $\widetilde{\chi}$ correctly describes excitonic 
effects in the polarization function at the BSE level~\cite{rmp}.
Even if $\tGa$ is an highly non--local, three--point function, 
it has been recently shown that, as long as we are interested in the 
two--point polarization function $\tilde{\chi}$,
Eq.\,(\ref{chitilde}) can be cast in terms of 
the two--point exchange--correlation kernel $f_{xc}$
of TDDFT~\cite{noi,francesco}:
\begin{align}
\tilde{\chi}\(1,2\)=\chi_0\(1,2\)+\int d34\chi_0\(1,3\)
f_{xc}\(3,4\)\tilde{\chi}\(4,2\).
\label{tddft}
\end{align}
Here 
$\chi_0\(1,2\)=-iG_0\(1,2\)G_0\(2,1\)$ gives the DFT polarization function.
At this point  if we take the exchange--correlation potential 
corresponding to $f_{xc}$ as a local approximation to the self--energy,
then the vertex function can be easily contracted into a two--point
function:
$\tGa\(6,7;3\)\equiv\tGa_{loc}\(6,3\)\gd\(6,7\)$~\cite{mahan,carlolaf,rodolfo}, with
\begin{align}
\tGa_{loc}\(1,2\)=\[\gd\(1,2\)-\int\,d3\,f_{xc}\(1,3\)\chi_0\(3,2\)\]^{-1}.
\label{gammaloc}
\end{align}
Thus Eq.\,(\ref{dyson}) gives $\Sigma\(1,2\)=i W^{TDDFT}\(1^+,2\)G_0\(1,2\)$,
in terms of the TDDFT effective potential 
$W^{TDDFT}\(1,2\)=
 \int\,d3\,v(1,3)\[\gd\(3,2\)-\int\,d4\,\(v\(3,4\)+f_{xc}\(3,4\)\)\chi_0\(4,2\)\]^{-1}$.
From this self--energy the lifetime of a generic conduction
state $c$ with momentum $\kk$ is given by
\begin{multline}
\tau^{-1}_{c\kk}=-2\Omega^{-1}\sum_{\GG_1,\GG_2}\sum_{\qq,c'}
\rho_{cc'}\(\kk\qq\GG_1\)\rho^*_{cc'}\(\kk\qq\GG_2\)\\
Im\[W^{TDDFT}_{\GG_1\GG_2}\(\qq,\gee_{c\kk}-\gee_{c' \kk-\qq}\)\],
\label{lifeloc}
\end{multline}
with $\rho_{nn'}\(\kk\qq\GG\)=\la
n\kk|e^{i\(\qq+\GG\)\cdot\rr}|n'\kk-\qq\ra$, $\GG$ a reciprocal space 
vector and $\Omega$ the crystal volume.
Different expression for $\tGa_{loc}$, based either on local--field
factor of the homogeneous electron gas~\cite{mahan} or on 
time--dependent local--density approximation\,(TDLDA)~\cite{carlolaf,rodolfo} have shown
that inclusion of local vertex corrections in both 
$\Sigma$ and $\tilde{\chi}$ almost cancel out, i.e., $\tGa_{loc}$ in $\Sigma$
undresses the exchange--correlation effects included in the polarization function $\tilde{\chi}$.
However such approximations for $f_{xc}$ produce optical spectra very 
similar to RPA, in disagreement with experiments.
This important drawback of a TDLDA $f_{xc}$ has been recently
related to the {\it long--range} nature of the kernel,
$f_{xc}\(\rr,\rr';\go\)\sim -\ga\(\go\)/|\rr-\rr'|$ that partially counteracts
the repulsive Hartree contribution~\cite{noi,francesco}.
The stronger the electron-hole effects are, the larger is the correction embodied in
$\ga$.
In the case of wide--gap insulators like LiF there is a large region of
frequencies and transfer momenta $\qq$ where 
$f_{xc}$ is stronger than the Hartree term (i.e. $\ga>1$).
This leads to unphysical linewidths: 
for a large energy range $Im\(\Sigma\)$,
and hence $\tau^{-1}$, has a wrong sign\,!
This result is  visualized by noticing that with respect to a $G_0W_0$ calculation
a change of  sign of $\tau^{-1}$
is controlled by the $sign\(v+f_{xc}\)$, that is
proportional to $\(1-\ga\(\go\)\)$. 
A similar result was obtained  in Ref.~\cite{carlolaf} looking at
the high $\qq$ limit of the TDLDA kernel that goes as $f_{xc}\sim q^2$.
The reason for this important failure of a two--point vertex function
is connected to the imposed reduction of the non--locality 
from the original, three--point vertex function.
In physical terms $\tGa_{loc}$ overestimates the intensity of 
the vertex correction because  
two incoming particles (entering in $1$ and $2$
in the exact  vertex function  $\tGa$) are supposed to
coexist at the same time--space point.
To overcome this difficulty we decided to release the constrain on the
spatial locality and define a TDDFT vertex function $\tGa_{TDDFT}\(1,2;3\)$ such that, for
a given $f_{xc}\(1,2\)$, $\tGa_{TDDFT}$ is
consistent with Eqs.\,(\ref{chitilde}--\ref{tddft}).
To this end we recall that in Ref.~\cite{noi} 
we derived a diagrammatic expression for $f_{xc}$ in terms of
the screened coulomb potential $W_0$, that to first order reads:
$f_{xc}=\chi_0^{-1}\tChi^{\(1\)}\chi_0^{-1}$,
with $\tChi^{\(1\)}$  the first order expansion of Eq.\,(\ref{chitilde})
in $W_{0}$~\cite{tokatly}. 
From this $f_{xc}$ we get an approximation for the 
vertex function $\tGa^{\(1\)}_{TDDFT}\(1,2;3\)$ 
imposing that once plugged in Eq.\,(\ref{chitilde}) it reproduces
Eq.\,(\ref{tddft}) for $\tilde{\chi}$.
By inspecting Eqs.\,(\ref{chitilde}--\ref{tddft}) we obtain
\begin{multline}
\tGa^{\(1\)}_{TDDFT}\(1,2;3\)\equiv\gd\(1,2\)\gd\(2,3\)+\\
i W_0\(1,2\) \int\,d4\,G_0\(1,4\)G_0\(4,2\)\tGa_{loc}\(4,3\),
\label{vertex}
\end{multline}
It is crucial to observe that $\tGa^{\(1\)}_{TDDFT}$ {\it is not} a first order vertex, as
$\tGa_{loc}$ sums an infinite number of diagrams.
Eq.\,(\ref{vertex}) can be easily generalized to give higher order approximations for
$\tGa$, consistent with the high order corrections to $f_{xc}$ of Ref.~\cite{noi}.
As it is commonly done we
neglect dynamical effects in the BSE~\cite{dbse}, i.e. we
assume $W_0\(1,2\)\approx W_0\(\rr_1,\rr_2;\go=0\)$ in Eq.\,(\ref{vertex}).
This approximation is motivated in the present case, as
we are interested in the low--energy electronic linewidths 
neglecting self--consistency effects.
We have verified numerically that for LiF, Si, diamond and SiO$_2$ 
$\tGa^{\(1\)}_{TDDFT}\(1,2;3\)$ is an excellent  approximation
to the ``true'' BSE vertex function $\tGa$. 

Now we can study the quasielectron lifetime in this approximation
for the vertex function and for the electronic self--energy.
To do so, we use as above the ``on mass--shell'' approximation, i.e.,  
the lifetime is given by the imaginary part of $\Sigma=G_0 W\tGa^{\(1\)}_{TDDFT}$ 
evaluated at the DFT energies:
\begin{align}
\tau^{-1}_{c\kk}= - Im\[\la c\kk|\Sigma\(\rr,\rr';\gee_{ c\kk}\)+
 \bar{\Sigma}\(\rr,\rr';\gee_{c\kk}\)|c\kk\ra\].
\end{align}
Here the linewidths are computed as an average of the ``left'' and ``right''
self--energies, $\Sigma$ and $\bar{\Sigma}$~\cite{strinati} in order to
restore the proper $\rr,\rr'$ symmetry of the  self--energy.
Using Eq.\,(\ref{vertex}) $\tau^{-1}_{c\kk}$ can be simplified by 
performing the energy integration in the complex plane and exploiting the 
pole structure of $\tGa^{\(1\)}_{TDDFT}$:
\begin{multline}
\tau^{-1}_{c\kk}=\tau^{-1}_{c\kk,0}\\-2\Omega^{-1}\sum_{\GG_1,\GG_2}\sum_{\qq,c'}
Im\[W^{TDDFT}_{\GG_1\GG_2}\(\qq,\gee_{c\kk}-\gee_{c' \kk-\qq}\)\]\\
Re\[\(\Gamma^{cv}_{cc'}\(\kk\qq\GG_1\)+\Gamma^{vc}_{cc'}\(\kk\qq\GG_1\)\)
   \rho^{*}_{cc'}\(\kk\qq\GG_2\)\],
\label{life1}
\end{multline}
where $\tau^{-1}_{c\kk,0}$ corresponds to the standard $G_0W$ approximation, and
\begin{multline}
\Gamma^{cv}_{cc'}\(\kk\qq\GG\)=\Omega^{-1}\sum_{\GG_1\GG_2\QQ}\sum_{c_2v_2}
\gr_{cc_2}\(\kk\qq\GG_1\)\[W_{0}\(\QQ\)\]_{\GG_1\GG_2}\\ \gr^*_{c'v_2}\(\kk-\qq\QQ\GG_2\)
\gr_{c_2v_2}\(\kk-\QQ\qq\GG\)\\
 \(\gee_{c\kk}-\gee_{c'\kk-\qq}-\gee_{c_2\kk-\QQ}+\gee_{v_2\kk-\qq-\QQ}\)^{-1},
\label{life2}
\end{multline}
\begin{multline}
\Gamma^{vc}_{cc'}\(\kk\qq\GG\)=\Omega^{-1}\sum_{\GG_1\GG_2\QQ}\sum_{c_2v_2}
\gr_{cv_2}\(\kk\qq\GG_1\)\[W_{0}\(\QQ\)\]_{\GG_1\GG_2}\\ \gr^*_{c'c_2}\(\kk-\qq\QQ\GG_2\)
\gr_{v_2c_2}\(\kk-\QQ\qq\GG\)\\
 \(\gee_{c\kk}-\gee_{c'\kk-\qq}-\gee_{v_2\kk-\QQ}+\gee_{c_2\kk-\qq-\QQ}\)^{-1}.
\label{life3}
\end{multline}
Eq.\,(\ref{life1}) constitutes the main basic result of this communication,
and can be easily extended to the quasihole linewidths.
Eq.\,(\ref{lifeloc})  must be compared with Eq.\,(\ref{life1}).
In the case of weakly
interacting systems the two equations with a TDLDA $f_{xc}$
give very similar quasiparticle corrections to the gap and
electron linewidths~\cite{mahan,carlolaf,rodolfo}. 
But, as short--range correlations become important
Eq.\,(\ref{lifeloc}) tends to give non--sensible results (negative linewidths) 
because of the wrong sign of $W^{TDDFT}$.
In Eq.\,(\ref{life1}), instead, the term $\(\Gamma^{cv}_{cc'}+\Gamma^{vc}_{cc'}\)$
reflects the spatial non--locality of $\tGa^{\(1\)}_{TDDFT}$
strongly reducing the weight of $W^{TDDFT}$.
Consequently the final expression for $\tau^{-1}_{c\kk}$ is given by
$\tau^{-1}_{c\kk,0}$ plus a small vertex correction that does not
change appreciably the results of a simpler $G_0W$ calculation.
The fundamental practical result of this work corresponds
to the solution of Eq.\,(\ref{life1}) for LiF, shown in Fig.~\ref{fig1}.
The overall effect of excitons in the
linewidths is huge~\cite{note}: 
the linewidths up to 3\,eV above
the forbidden region display a linear dependence with energy
while the RPA are almost zero because of the slow rise
of the RPA loss function  (see Fig.\ref{fig2}). 
A similar energy dependence has been observed in  highly correlated
materials~\cite{smith}. Instead the present 
linear dependence of the linewidths is due to
the combination of an almost constant density--of--states
close to the conduction band minimum and to
a ``step--like'' energy dependence of loss function (see Fig.\ref{fig2}).
Furthermore, the quasiparticle
linewidths are not exactly zero in a small energy window of $0.5$\,eV in
the forbidden region. This effect can be traced back to the
excitonic--induced transfer 
of oscillator strength in the dynamical dielectric
function {\it below} the gap. This result is consistent with the fact
that exciton dynamics is dictated by vertex--correction to the
self--energy, therefore an interpretation of the quasiparticle scattering
based only on independent--particle processes losses meaning.
The results of the present work allow for the systematic analysis 
of the role of excitons in quasiparticle excitations and response
functions of extended and low dimensional systems, where the standard 
$G_0W_0$ approximation fails~\cite{gwrep,rmp}.

This work was supported by the European Community Research Training
Network NANOPHASE (HPRN-CT-2000-00167) and Network of Excellence
NANOQUANTA (NOE 500198-2).
We acknowledge fruitful discussions with Lucia Reining and the precious 
support and suggestions of Rodolfo Del Sole in deriving Eq.\,(\ref{vertex}).

%%%%%%%%%%%%%%%%%%%%%% REFERENCES %%%%%%%%%%%%%%%%%%%%%%%%%


\begin{thebibliography}{100}
\bibitem{metals}
 In metals, instead, the deviations from the Fermi--liquid square root 
 dependence of the lifetime
 ($\tau \propto (E-E_{F})^{-2}$) close to the Fermi level
 were considered a fingerprint of strong--correlation effects~\cite{smith}.
\bibitem{details} 
 Even if the solution of the BSE equation would be possible there are two
 main points in favor of a calculation based on $\fxc$: i) all the
 components of the dielectric function are obtained in one run,
 ii) it is feasible to perform the 
 extrapolation to zero-broadening $\eta$ in the polarization
 to compute the lifetime (note that for low--energy states
 the difference between a calculation
 with $\eta=100$\,meV and $\eta=10$\,meV can be up to $\sim$400\%).
 In order to do a sensible BSE calculation for small broadening
 it is necessary to use a large number of $\kk$-points making the BSE 
 diagonalization unpractical. This is not the case for the TDDFT approach.
\bibitem{note} When comparing the results of  Fig.~\ref{fig1} with the
 experiment it must be noticed that the ``on mass--shell'' neglects
 the QP renormalization factors, and it commonly overestimates
 the electronic linewidths.
\end{thebibliography}
\end{document}